\documentclass[aps,prl,twocolumn,nopacs,amsmath,amssymb]{revtex4-1}
\usepackage{dcolumn}
\usepackage{bm}
\usepackage{subfigure}
\usepackage{graphicx}
\usepackage{hyperref}
\usepackage{lipsum}
\usepackage{url}
\usepackage[usenames,dvipsnames,svgnames]{xcolor}

\renewcommand{\section}[1]{{\par\it #1.---}\ignorespaces}

\begin{document}
	
	\makeatletter
	\newcommand*{\balancecolsandclearpage}{%
		\close@column@grid
		\clearpage
		\twocolumngrid
	}
	\makeatother

	\title{Probing unconventional superconductivity in proximitized graphene by impurity scattering}
	\author{Oladunjoye A. Awoga}
	\author{Annica M. Black-Schaffer}
	\affiliation{Department of Physics and Astronomy, Uppsala University, Box 516, S-751 20 Uppsala, Sweden}
	\begin{abstract}
		We demonstrate how potential impurities are a very powerful tool for determining the pairing symmetry in graphene proximity-coupled to a spin-singlet superconductor. All $d$-wave states are characterized by subgap resonances, with spatial patterns clearly distinguishing between nodal and chiral $d$-wave symmetry, while $s$-wave states have no subgap resonances. We also find strong supergap impurity resonances associated with the normal state Dirac point. Sub- and supergap resonances only interact at very low doping levels, then causing suppression of the supergap resonances.
	\end{abstract}
	\maketitle
	
	Superconductivity in graphene~\cite{novoselov2004electric}, a honeycomb lattice Dirac material~\cite{WehlingBlack2014}, has attracted significant interest. While superconductivity in pristine graphene has so far been experimentally elusive, it has been achieved through alkali-metal deposition \cite{mcchesney2010extended,ludbrook2015evidence,chapman2016superconductivity, ichinokura2016superconducting} and by proximity to an external superconductor (SC). In the latter case both conventional spin-singlet $s$-wave SCs \cite{heersche2007bipolar,tonnoir2013induced,natterer2016scanning,bretheau2017tunnelling} and, recently, a spin-singlet $d$-wave cuprate SC \cite{di2017corrigendum} have been used, making it important to consider different pairing symmetries.
	
	The superconducting pairing symmetry is crucial as it is the key for many properties in the superconducting state. However, the induced pairing symmetry can be unknown even in proximitized systems, due to interface roughness or special properties of the material \cite{BlackSchaffer2013proximity}. Graphene offers challenges particularly due to its sixfold rotationally symmetric lattice, which is not naturally compatible with the fourfold $d$-wave states \cite{BlackSchafferandHonerkamp2014}.
	This results in the chiral $d_{x^2-y^2}\pm id_{xy}$-wave combination having the lowest intrinsic energy among the $d$-wave states \cite{Black-Schaffer07,Jiang2008andreev, nandkishore2012chiral,Wang11,Kiesel12}. This is a fully gapped sixfold symmetric topological state, which hosts two chiral edge states. In contrast, the $d_{x^2-y^2}$ (or $d_{xy}$) state has a nodal V-shaped energy spectrum.
	
	Recent scanning tunneling spectroscopy (STS) experiments \cite{di2017corrigendum} on graphene proximitized by the $d$-wave cuprate ${\mathrm{Pr}}_{2\ensuremath{-}x}{\mathrm{Ce}}_{x}{\mathrm{CuO}}_{4}$ \cite{Qazilbash2003Point, Dagan2007, Kalcheim2012evidence} have detected both V-shaped differential conductance and zero-energy conductance peaks. These were proposed to be due to faceting of the cuprate surface, in combination with an effective spin-singlet $p$-wave state in graphene. Notably, a nodal $d$-wave state (symmetry with respect to the center of the Brillouin zone (BZ)) results in effective $p$-wave symmetry at the Fermi surfaces around the $K,K'$ BZ corners \cite{Linder2009}. However, the low doping levels in graphene results in very small superconducting gaps \cite{WehlingBlack2014, BlackSchafferandHonerkamp2014}, such that STS can lack the resolution to clearly resolve between gapped $s$- or chiral $d$-wave states and nodal $d$-wave states. 
	
	Modifications of the local density of states (LDOS) by a single potential impurity offers a tantalizing opportunity to more accurately determine the superconducting pairing symmetry. While potential impurities cannot induce subgap states in conventional $s$-wave SCs, as stated in Anderson's theorem \cite{anderson1959dirty}, higher angular momentum  and unconventional SCs usually host distinct sets of subgap impurity states \cite{Balatskyet.alRMP2006,tsai2009impurity}. This has successfully been used in STS experiments to identify the pairing symmetry of unconventional SCs  \cite{Yazdani1999impurity,pan2000imaging,Fischer2007QPI,zhou2013visualizing}, including using Fourier transformation for the quasiparticle interference (QPI) \cite{hoffman2002imaging,Wang2003Quasiparticle,mcelroy2003relating,bena2005quasiparticle,rutter2007scattering,BrihuegaQPI2008,allan2013imaging}. Moreover, potential impurities in normal-state graphene give universal resonance peaks near the Dirac point \cite{Balatskyet.alRMP2006,WehlingBlack2014,wehling2007local,Ugeda2010missing,lado2016unconventional}. Potential impurities in superconducting graphene will thus likely not only give rise to rich physics, but importantly provide a route to determine the pairing symmetry.  
	
	In this work we show that a potential impurity in graphene in proximity to any spin-singlet SC gives rise to several impurity resonances that uniquely determine the pairing symmetry. In particular, any $d$-wave state is characterized by  subgap resonances, with spatial and QPI patterns clearly differentiating between nodal and chiral states. 
	In contrast, $s$-wave states have no subgap resonances. Constant and extended $s$-wave symmetry is however still easily distinguishable since the latter is nearly gapless at low doping levels. 
	We also find that superconducting graphene always hosts supergap impurity resonances, associated with the normal state Dirac point. Subgap and supergap impurity resonances do not interact as long as the normal state Dirac point is well separated in energy from the superconducting gap edge due to doping. However, at very low doping levels the supergap resonance is strongly suppressed. 
	Our results are directly relevant to experiments on proximity-induced superconductivity in graphene~\cite{tonnoir2013induced, natterer2016scanning, bretheau2017tunnelling, di2017corrigendum} and show how impurities are exquisitely suitable to probe the pairing symmetry.

	\section{Model}
	\begin{figure}[htb]
		\centering
		\includegraphics[scale=0.44]{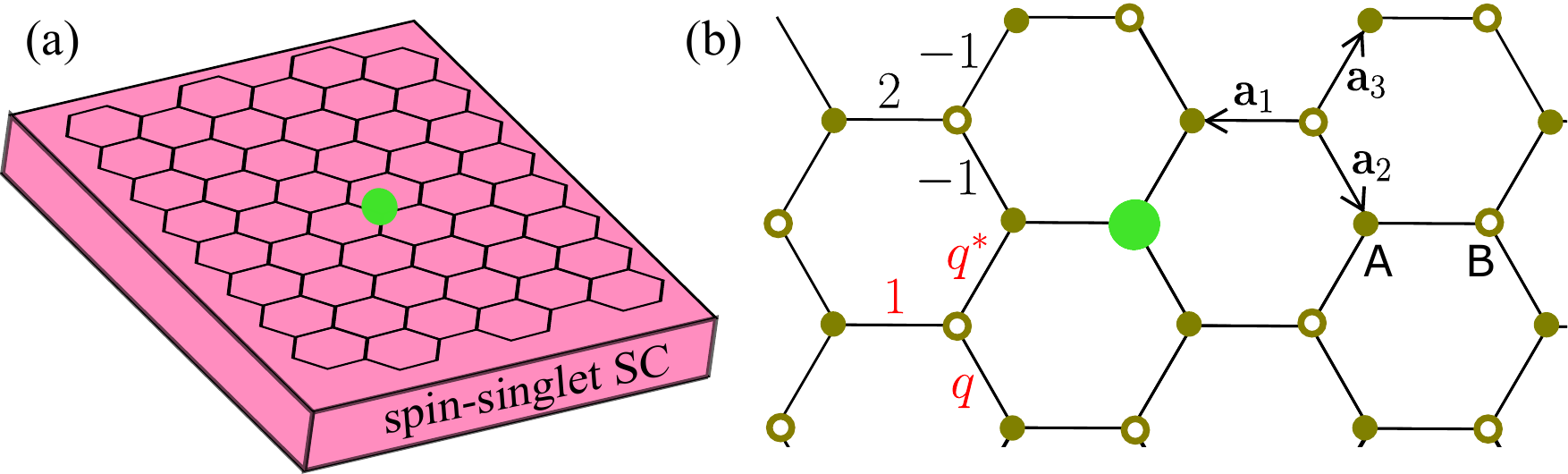}
		\caption[Schematic]{(a) Schematic setup of graphene in proximity to a spin-singlet SC with a single impurity (green disk). (b) Honeycomb lattice with sublattice A (filled) and B (open), NN bonds directions $\mathbf{a_\lambda}$, and (unrenormalized) $d_{x^2-y^2}$- (black) and $d_{x^2-y^2}+id_{xy}$-waves (red) bond pairing symmetry factors, with $q = e^{i2\pi/3}$.}
		\label{fig:ProxGraphene}
	\end{figure}
	When graphene is placed on a SC, see Fig.~\ref{fig:ProxGraphene}(a), Cooper pairs are injected into the graphene layer through Andreev processes~\cite{andreev1964thermal,beenakker2008colloquium}.  Here we consider a spin-singlet SC substrate but allow for all $s$- and $d$-wave spatial symmetries, to cover both conventional and cuprate SCs as substrates. In recent STS measurements of the graphene layer \cite{di2017corrigendum}, the substrate SC was found to not influence the results to any significant degree. We can thus safely consider only the graphene layer with proximity-induced superconductivity, which also makes our results applicable to superconductivity induced by alkali-metal doping. In total, the effective mean-field Hamiltonian $\mathcal{H}= \mathcal{H}_0 + \mathcal{H}_\Delta$ is given by \cite{Black-Schaffer07,Lothman14,BlackSchafferandHonerkamp2014,Awoga2017DomainWall}
	%
	\begin{align}\label{eq:Hamiltmf} 
		\mathcal{H}_0&=-t\sum_{i,\lambda,\sigma}(a_{i\sigma}^\dag b_{i+\mathbf{a}_\lambda\sigma}+\textrm{H.c.})
		+ \mu\sum_{i,\sigma}(a_{i\sigma}^\dag a_{i\sigma}+b_{i\sigma}^\dag b_{i\sigma}) \nonumber \\                                                                      
		\mathcal{H}_\Delta&= \sum_{i,\lambda} \lbrace\Delta_\lambda (a_{i\uparrow}^\dag    
		b_{i+\textbf{a}_\lambda\downarrow}^\dag-a_{i\downarrow}^\dag b_{i+\textbf{a}_\lambda\uparrow}^\dag)  
		+ \textrm{H.c.}\rbrace 
		\nonumber \\
		&+\Delta_\textrm{on} \sum_{i} \lbrace (a_{i\uparrow}^\dag    
		a_{i\downarrow}^\dag+b_{i\downarrow}^\dag b_{i\uparrow}^\dag)  
		+ \textrm{H.c.}\rbrace .
	\end{align}
	Here $a_{{i}\sigma}^\dag (b_{{i}\sigma}^\dag)$ creates an electron with spin $\sigma$ at site $i$ in sublattice A (B) and $\textbf{a}_\lambda$ with $\lambda = 1,2,3$ being the nearest neighbor bond (NN) directions, see Fig.~\ref{fig:ProxGraphene}(b).
	$\mathcal{H}_0$ is the normal state Hamiltonian with NN hopping $t$ and effective chemical potential $\mu$ due the substrate and possibly gating. 
	$\mathcal{H}_\Delta$ gives the proximity-induced spin-singlet superconductivity, including both pairing on NN bonds $\Delta_\lambda$ and on-site pairing $\Delta_{\rm on}$. This form captures all realistic spin-singlet pairing symmetries allowed by the $D_{6h}$ point group of graphene \cite{BlackSchafferandHonerkamp2014}, namely: conventional $s$-wave symmetry from $\Delta_{\rm on}$ and extended $s$-wave ($s_\text{ex}$), $d_{x^2-y^2}$, and $d_{xy}$-wave symmetries generated by $\Delta_{\lambda}$. The basis functions for the latter states over the three NN bonds are $\boldsymbol{\Delta}_{s_\text{ex}} =\frac{1}{\sqrt{3}}(1,1,1)$, $\boldsymbol{\Delta}_{d_{x^2-y^2}} = \frac{1}{\sqrt{6}}(2,-1,-1)$, and $\boldsymbol{\Delta}_{d_{xy}} = \frac{1}{\sqrt{2}}(0,1,-1)$. 
	Moreover, the two $d$-wave solutions belong to the same irreducible representation of the point group and are thus allowed to mix. In particular, the chiral combination $\boldsymbol{\Delta}_{d_{x^2-y^2}\pm id_{xy}}= \frac{1}{\sqrt{3}}(1,e^{\pm i\frac{2\pi}{3}}, e^{\pm i\frac{4\pi}{3}})$ is a fully gapped time-reversal symmetry breaking state that has been shown to have the lowest intrinsic energy \cite{Black-Schaffer07, nandkishore2012chiral, Wang11, Kiesel12, BlackSchafferandHonerkamp2014}, since either $d$-wave state is nodal. The restriction to NN pairing is appropriate for any cuprate substrate, but longer-range pairing has previously been shown to not significantly change any properties \cite{BlackSchaffer12PRL,Kiesel12}
	
	We use $\boldsymbol{\Delta}=\Delta_0  {\boldsymbol{\Delta}}_r$, where ${\boldsymbol{\Delta}}_r$ are the basis functions, and measure the strength of the pairing by the magnitude $\Delta_0$ for all symmetries, but note that the superconducting gap edge energy also depends on the symmetry and doping. For large doping levels relative to the superconducting gap, such that $\beta =|\mu|/\Delta_0 \gg 1$ \cite{GraphImpRatio}, the gap edge is always well separated from the normal state Dirac point ($D_g$) found at $E = \mu$. It gives fully gapped states at the Fermi level for the $s$-, $s_{\rm ex}$-, and chiral $d$-wave symmetry, while the $d_{x^2-y^2}$- and $d_{xy}$-wave states are nodal superconductors, see figure in Supplementary Material (SM) \cite{SMGraphImp}. The two latter systems are thus double Dirac point systems, with Dirac points generated both from  $d$-wave superconductivity and normal state graphene. At $\beta \sim 1$ the superconducting coherence peaks fills $D_g$ with additional states such that it is washed out. The only exception is the $s_{\rm ex}$-wave state as it becomes a hidden order with no superconducting gap as $\mu$ decreases \cite{Uchoa2007}. 
	
	To investigate the influence of impurities, we add a single potential impurity of strength $U$ to site $m$ using $\mathcal{H}_\text{imp}=U\sum_\sigma c_{m\sigma}^\dag c_{m\sigma}$. We set $\mu> 0, U>0$ which adds positively to the total on-site energy of site $m$ (for a discussion different signs, see SM \cite{SMGraphImp}).
	The full Hamiltonian $\mathcal{H}_{\rm tot}=\mathcal{H}_0+\mathcal{H}_\Delta+\mathcal{H}_\text{imp}$ is solved within the Bogoliubov-de-Gennes framework using a Chebyshev polynomial expansion approach~\cite{weiss2006kernel,covaci2010,Covaci2011Proximity} implemented in the TBTK code package~\cite{kristofer_bjornson_2017_997267,kristofer_bjornson_2018_1149702}. The Chebychev expansion allows us to study considerably larger lattices, such that edge states have no influence on the results, at significantly reduced computational cost compared to exact diagonalization. We can also safely ignore the influence of the impurity on the order parameter itself in order to save computational cost. While self-consistent calculations will find a suppression of $\Delta_0$ and possibly a slight change of symmetry extremely localized to the impurity \cite{Flatte1997local, Salkola1997spectral, bjornson2016piphase, Lothman14}, it will not alter the existence of subgap impurity resonances \cite{Lothman14, mashkoori2017impurity} nor notably influence their extended spatial extent.

	\section{Supergap resonances}
	For finite impurity strength $U$, the LDOS is suppressed at the impurity site, see SM~\cite{SMGraphImp}. The largest LDOS effects of the impurity is therefore seen on its NN sites. We therefore first discuss the LDOS at impurity NN sites and later turn to the long-range spatial properties.
	\begin{figure}[htb]
		\centering
		\includegraphics[scale=0.31]{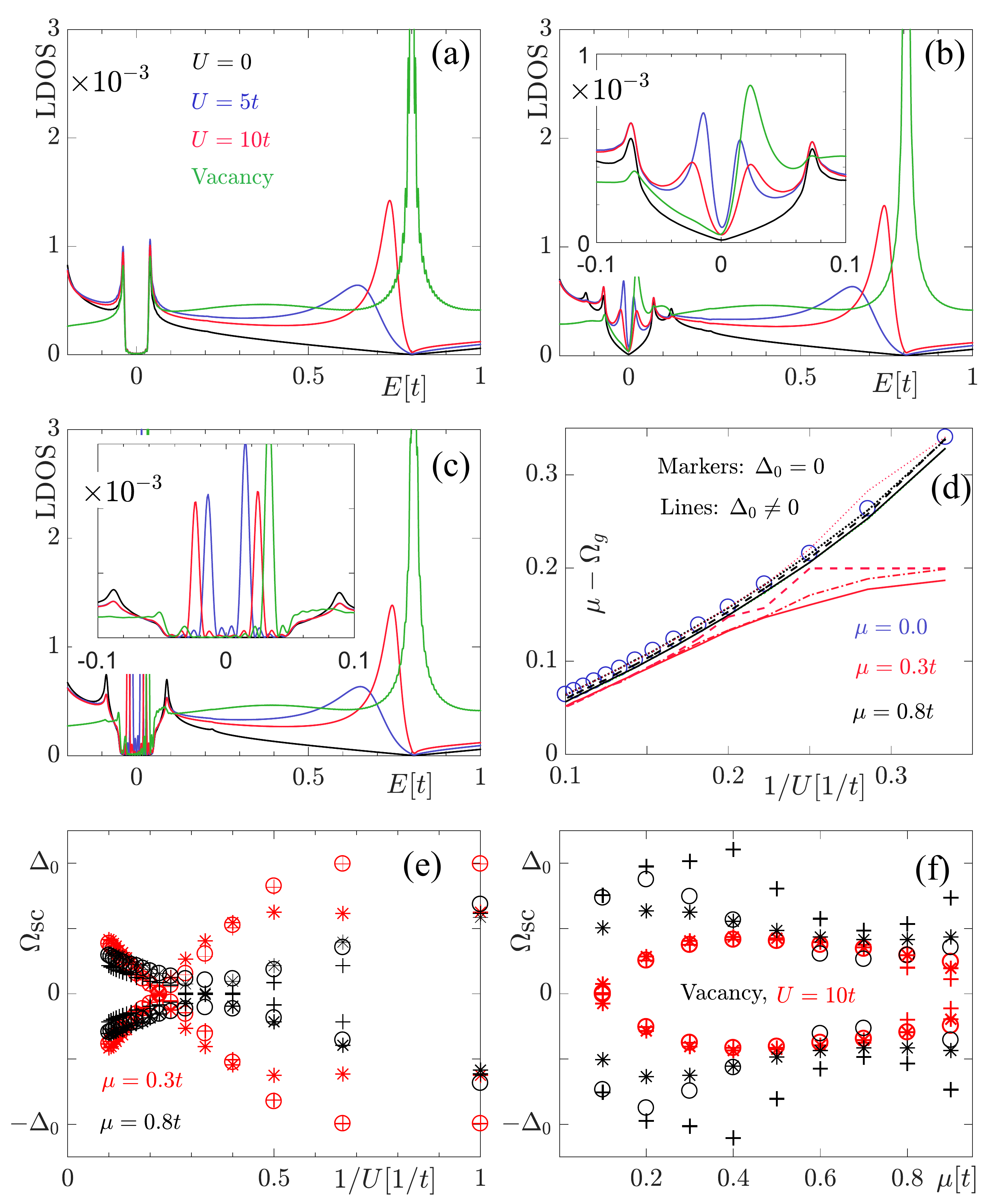}
		\caption[LDOS]{LDOS at impurity NN sites for different $U$ in the $s_\textrm{ex}$ (a), $d_{x^2-y^2}$ (b) and $d_{x^2-y^2}+id_{xy}$ (c) states for $\mu=0.8t$, $\beta=10$. Insets show a zoom-in at low energies. 
			(d) Supergap peak energy $\Omega_{g}$ shifted by $\mu$ as a function of $1/U$ for non-superconducting ($\circ$), $s_\textrm{ex}$ (dotted), $s_\textrm{on}$ (dashed), $d_{x^2-y^2}$ (solid), and $d_{x^2-y^2}+id_{xy}$ (dashed-dotted) states.  (e, f) Subgap peak energies $\pm\Omega_{sc}$ for $d_{xy}$ ($+$), $d_{x^2-y^2}$ ($\circ$), and $d_{x^2-y^2}+id_{xy}$ ($\ast$) states as a function of $1/U$ and $\mu$. Here $\Delta_0 = 0.08t$.
		}
		\label{fig:LDOS}
	\end{figure}
	In Figs.~\ref{fig:LDOS}(a-c) we show the LDOS at impurity NN sites for $s_{\rm ex}$, $d_{x^2-y2}$, and $d_{x^2-y^2}+id_{xy}$ superconducting symmetries, respectively, in the limit of large doping (here $\beta = 10$) and increasing $U$ up to the unitary scattering limit $U\rightarrow \infty$, modeling a vacancy. The first striking result is that we see the exact same strong supergap impurity resonance close to $D_g$ (at $E =0.8t$) in all cases. The peak energy, $\Omega_g$, varies with $U$ and is the same as in the normal state (see SM \cite{SMGraphImp}). Impurity resonances is a well-known characteristic of Dirac materials, with peak energy universal scaling as $E_{\rm peak} \sim 1/U$ towards the Dirac point for normal state materials \cite{WehlingBlack2014}. The strong supergap resonances in all types of superconducting graphene is thus due to the normal state graphene Dirac point, $D_g$. 
	
	In Fig.~\ref{fig:LDOS}(d) we plot $\Omega_g$ as a function of $1/U$ for different $\mu$ and all superconducting states. Rescaling the energy with $\mu$, we find the peak energy for the non-superconducting case (circles) and all superconducting states at large $\beta$ (black lines) collapsing to a single, approximately linear, curve.
	This is due to doping simply shifting all energy levels, Dirac point and impurity resonances, by $\mu$. 
	For smaller $\beta$ (red lines), we start to see interference between the $D_g$ resonance and that of the superconducting gap, at first for lower values of $U$ where the $D_g$ impurity resonance is closer to the superconducting gap region.  As $\beta$ is even further reduced, the supergap resonance is strongly suppressed since $D_{g}$ is washed out by the superconducting coherence peaks, see SM~\cite{SMGraphImp}. 
	The extended $s_\text{ex}$ state (dotted lines) is the only exception. Here the $D_g$ impurity resonance does not change with $\mu$ since its superconducting gap disappears linearly with decreasing $\mu$.
	Remarkably, this means that the normal state Dirac point $D_g$, hitherto seemingly ignored \cite{wehling2008local,pellegrino2010pairing}, plays an important role for impurity physics also in superconducting graphene. Notably, the supergap peak can serve as an important experimental reference.
	
	\section{Subgap resonances}
	Turning our attention to the subgap resonance spectrum, we see in Fig.~\ref{fig:LDOS}(a) that the $s_\text{ex}$ state is fully gapped even with an impurity present, similar to the $s$-wave case and consistent with Anderson's theorem~\cite{anderson1959dirty}. Thus the resonance peak associated with $D_g$ is the only LDOS signature of impurities in (extended) $s$-wave superconducting graphene.
	In contrast, in Fig.~\ref{fig:LDOS}(b) we see that the nodal $d_{x^2-y^2}$-wave state has two spin degenerate virtual bound subgap resonances, with the $d_{xy}$ behaving similarly, see SM~\cite{SMGraphImp}. They are positioned symmetrically around zero energy, but with different heights due to particle-hole asymmetry in the normal state. The subgap resonances is consistent with a nodal $d$-wave SC being a Dirac material with the Dirac point fixed at $E = 0$ \cite{WehlingBlack2014} and have also been seen experimentally in the cuprate SCs \cite{Yazdani1999impurity,pan2000imaging}. Nodal $d$-wave superconducting graphene thus has a double set of Dirac-induced impurity resonances; at subgap energies from the $d$-wave Dirac point and at supergap energies from the normal state Dirac point, $D_g$. For large $\beta$ these are individually resolved but at $\beta\sim 1$ the supergap resonance is suppressed while the subgap peak largely remains, see SM~\cite{SMGraphImp}. This is thus a highly unusual situation of interaction between the two Dirac systems. 
	The fully gapped chiral $d_{x^2-y^2}+id_{xy}$-wave state in Fig.~\ref{fig:LDOS}(c) also hosts spin degenerate subgap resonances symmetric around $E=0$. These are real bound subgap states due to the full  gap, unlike the virtual states in the nodal $d$-wave state. These subgap resonances are consistent with results reported for defects in $d_{x^2-y^2}+id_{xy}$-wave superconducting graphene~\cite{Lothman14} and on a square lattice~\cite{mashkoori2017impurity}.
	
	The $d$-wave subgap resonance peaks $\Omega_\text{sc}$ changes with $U$, but the behavior is very different from that of the supergap resonance, see Figs.~\ref{fig:LDOS}(e,f). Although the behavior for all different $d$-wave states are qualitatively similar, there are differences and also a non-trivial dependence on $\mu$ not present for the supergap resonance. This includes the $U$ value needed for an accidental zero-energy crossing changing with doping.
	Based on these results, we conclude that the existence of subgap states directly identifies a $d$-wave state since all $s$-wave states have a clean superconducting gap.

	\section{Graphene on cuprate substrate}
	Above we established the existence of both supergap and subgap impurity resonances. We have however so far kept both $\mu$ and $\Delta_0$ relatively large, to clearly elucidate the different phenomena. To model a more realistic experimental regime, such as graphene on a cuprate substrate\cite{di2017corrigendum}, we now reduce $\mu$ and $\Delta_0$ by an order of magnitude but still keep $\beta = 10$ as in experiments. In this parameter regime we find that all superconducting states in the bulk actually show similar V-shaped LDOS with smearing akin to temperature effects, see SM~\cite{SMGraphImp}. This is in agreement with recent experiments~\cite{di2017corrigendum}, and makes it in fact impossible to distinguish between different superconducting symmetries using solely bulk LDOS measurements.
	\begin{figure}[htb]
		\centering
		\includegraphics[scale=0.31]{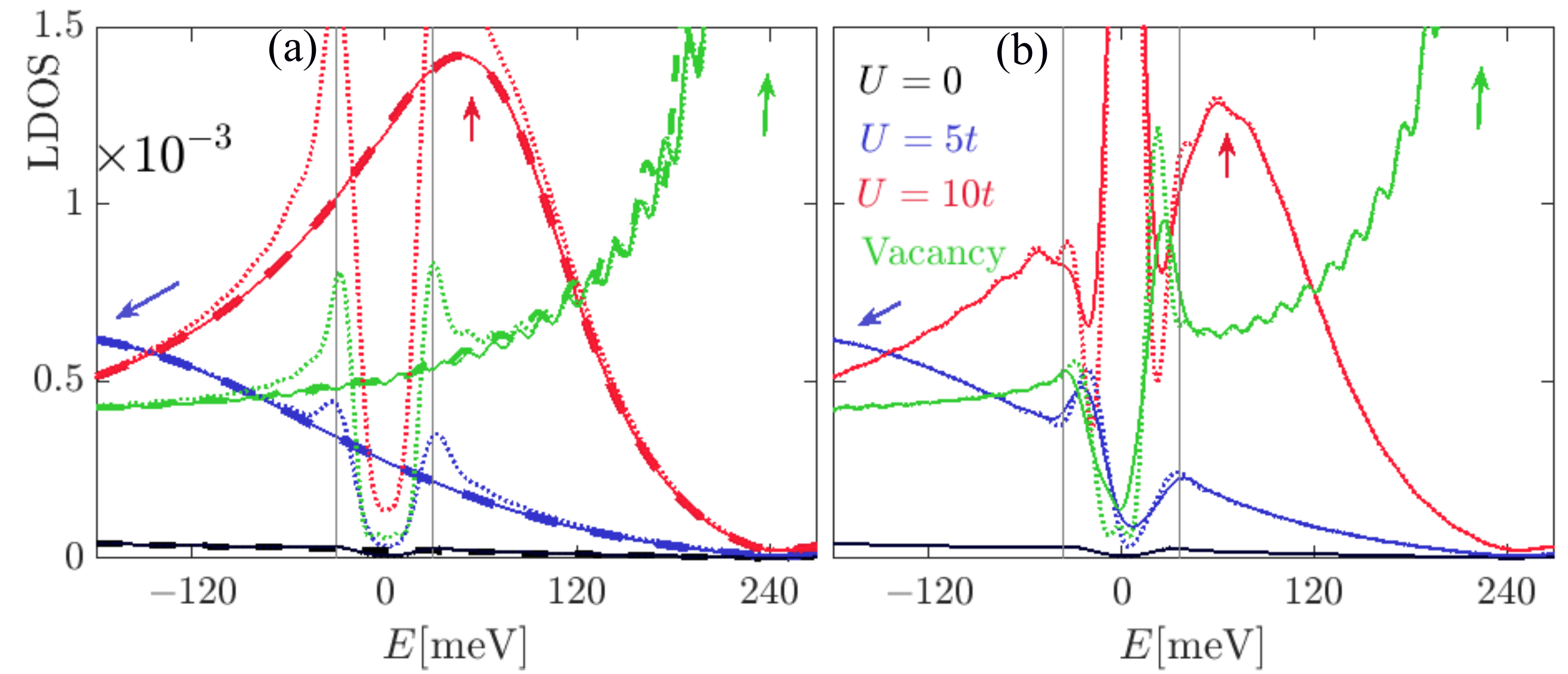}
		\caption[LDOS]{LDOS at impurity NN sites for different $U$ when $\mu=0.08t=240$~meV and $\Delta_0=0.008t=24$~meV (here $t = 3$~eV) giving $\beta=10$, for normal (dashed), $s_\textrm{ex}$ (solid) and $s_\textrm{on}$ (dotted) states (a) and $d_{x^2-y^2}$ (solid) and $d_{x^2-y^2}+id_{xy}$ (dotted) states (b). Arrows indicate resonance peaks due to $D_g$, vertical grey lines mark bulk gap edges.}
		\label{fig:LDOSMu008}
	\end{figure}
	However, we still see clear and distinguishing impurity effects as illustrated in Fig.~\ref{fig:LDOSMu008}. The supergap resonance due to $D_g$ is still present (arrow), but can now appear also at negative energies for lower $U$, since its energy is set only by the distance from $D_g$. 
	Most importantly, we see in Fig.~\ref{fig:LDOSMu008}(a) that all $s$-wave states still lack subgap states, despite the V-shaped LDOS. Also, the doping level is now low enough such that the $s_\text{ex}$ state (solid) show no gap, but instead looks completely non-superconducting. It is therefore possible to clearly differentiate between the $s$- and $s_{\rm ex}$ states by a bulk LDOS measurement in this experimentally relevant regime. 
	On the other hand, all $d$-waves states, see Fig.~\ref{fig:LDOSMu008}(b), still have strong subgap impurity resonances in a qualitatively similar manner as in Fig.~\ref{fig:LDOS}(b,c). Impurity effects in this experimentally relevant parameter regime is thus qualitatively the same as reported above. Further reducing the doping and superconducting gap does not change the results, see SM~\cite{SMGraphImp}.

	\section{Spatial pattern}
	The above results show how the existence of subgap resonances uniquely differentiate $d$-wave symmetries from any $s$-wave state. However, the difference between nodal and chiral $d$-wave superconductivity has so far been less clear, especially for smaller gaps where also a fully gapped state appears with an V-shaped bulk LDOS profile. We here study the full long-range spatial impurity modulation of the LDOS to possibly differentiate between the $d$-wave states. 
	
\begin{figure}[ht!]
		\centering
		\includegraphics[scale=0.40]{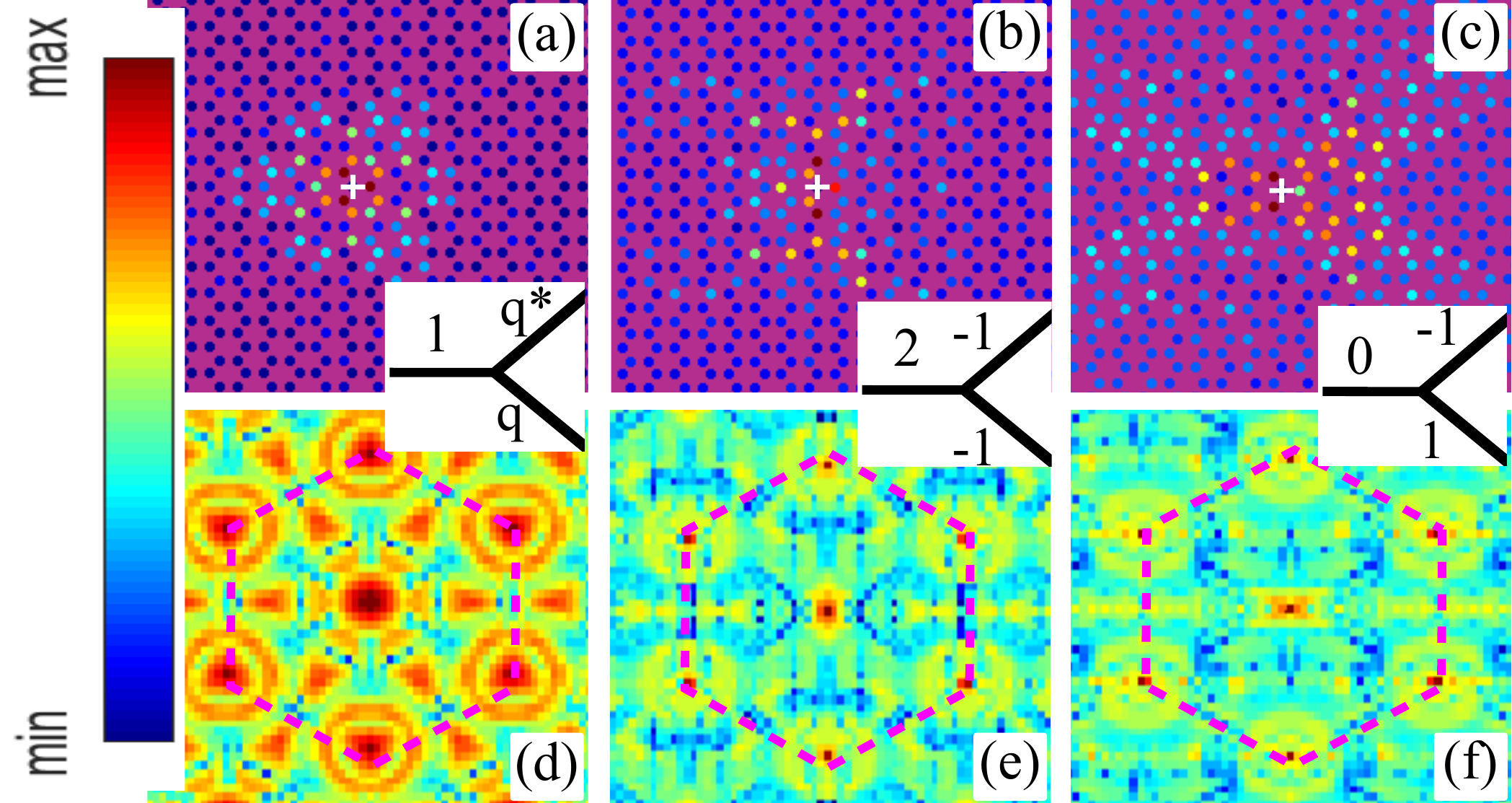}
		\caption[LDOSlattice]{Spatial variation of LDOS (top panels) and the logarithm of FTLDOS intensity (bottom panels) at negative subgap impurity resonance peak energy for $\mu=0.8t,\ \Delta_0=0.08t,\ U=10t$ for $d_{x^2-y^2}+id_{xy}$ (a,d), $d_{x^2-y^2}$ (b,e) and  $d_{xy}$ (c,f) states. White cross marks impurity site, purple hexagon depicts first BZ, insets show NN bond pairing symmetry factors. The LDOS (FTLDOS) intensity minimum is $0$ $(-1)$ while the maximum intensity is $1$ $(-0.3)$, with each figure normalized by its maximum intensity magnitude.}
		\label{fig:spLDOSEn}
	\end{figure}

	Impurity resonances in SCs give a distinct spatial pattern, not unlike Friedel oscillations in metals. For the supergap resonance we find the same spatial pattern in the superconducting states and the non-superconducting state, see SM~\cite{SMGraphImp}, which again illustrates that the supergap resonance is caused by the normal state Dirac point. 
	For the subgap resonances the spatial patterns however varies. In the fully gapped $d_{x^2-y^2}+id_{xy}$ state, see Fig.~\ref{fig:spLDOSEn}(a), the LDOS oscillations keep the full sixfold rotational symmetry of the lattice. This is to be expected since the state is both fully gapped and does not break the sixfold symmetry.
	In comparison, the $d_{x^2-y^2}$ and $d_{xy}$ states give distinct symmetry-breaking patterns, see Fig.~\ref{fig:spLDOSEn}(b) and (c), respectively. Roughly, the impurity states spread out in the real-space nodal regions of the order parameter, see insets showing the bond orders. There is also a slight preference for impurity states to follow the zigzag direction rotated $\tfrac{2\pi}{3}$ from the horizontal axis. Interestingly, domain wall states in chiral $d$-wave superconducting state in graphene has been found to have the same preference~\cite{Awoga2017DomainWall}.
	Depending on experimental setup, the orientation of the nodal $d$-wave state can change, and hence the exact spatial impurity patterns. Yet, the sixfold symmetry breaking is a clear signal distinguishing any nodal state from the chiral state. Thus, impurities can clearly distinguish between different $d$-wave symmetries, even when small full and nodal gaps both produce effective V-shaped bulk LDOS profiles.
	
	We can even further differentiate between the different $d$-wave states by studying the QPI pattern, given by the Fourier transformed LDOS (FTLDOS). 
	The QPI pattern of the chiral  $d$-wave state, see Fig.~\ref{fig:spLDOSEn}(d), has sixfold symmetry. There are pronounced rings at the $K, K'$ corners of the BZ and also a ring at $\mathbf{q}=0$, as well as structures symmetrically around the $M$-points. 
	The nodal $d$-waves, see Fig.~\ref{fig:spLDOSEn}(e,f), on the other hand distinctly break the sixfold symmetry, with rings that are no longer isotropic and an overall structure that has only twofold rotational symmetry. Furthermore, the feature at $\mathbf{q}=0$ is strongly suppressed compared to the chiral $d$-wave state. 
	Each of the features in the QPI corresponds to different scattering processes, but a detail analysis goes beyond the current objective. 
	In Fig.~\ref{fig:spLDOSEn} we plot the spatial (FT)LDOS for the negative energy impurity resonance, but the positive energy resonance gives similar information, see SM~\cite{SMGraphImp}. Moreover, $U$ and $\mu$ do not significantly influence the spatial behavior, although for $\beta \sim 1$ the rings at the BZ corners are suppressed, see SM~\cite{SMGraphImp}.

	In summary, we have demonstrated that potential impurities are an extremely powerful tool for identifying pairing symmetries in graphene proximity-coupled to any spin-singlet SC. Subgap states only exist for $d$-wave symmetries, but  distinct spatial patterns clearly differentiate chiral and nodal $d$-wave phases. We also find that superconducting graphene  hosts strong co-existing supergap impurity resonances due to the normal state Dirac point. 
	In combination this opens for very accurate determination of the pairing symmetry using STS measurements.

	We thank K.~Bj\"{o}rnson, C.~Triola and M.~Mashkoori for useful discussions.
	This work was supported by the Swedish Research Council (Vetenskapsr\aa det) Grant No.~621-2014-3721, the Swedish Foundation for Strategic Research (SSF), the G\"{o}ran Gustafsson Foundation, and the Wallenberg Academy Fellows program through the Knut and Alice Wallenberg Foundation.
	
	\bibliographystyle{apsrevmy}
	\bibliography{abbrRef,References}
	
	\balancecolsandclearpage
	\onecolumngrid
	\begin{center} {\large \bf Supplemental material}\end{center}
In this supplementary material we provide additional figures and accompanied discussion to further support the work and conclusions of the main text.

\subsection{Doping effect on bulk DOS}
\begin{figure}[ht!]
	\centering
	\includegraphics[scale=0.440]{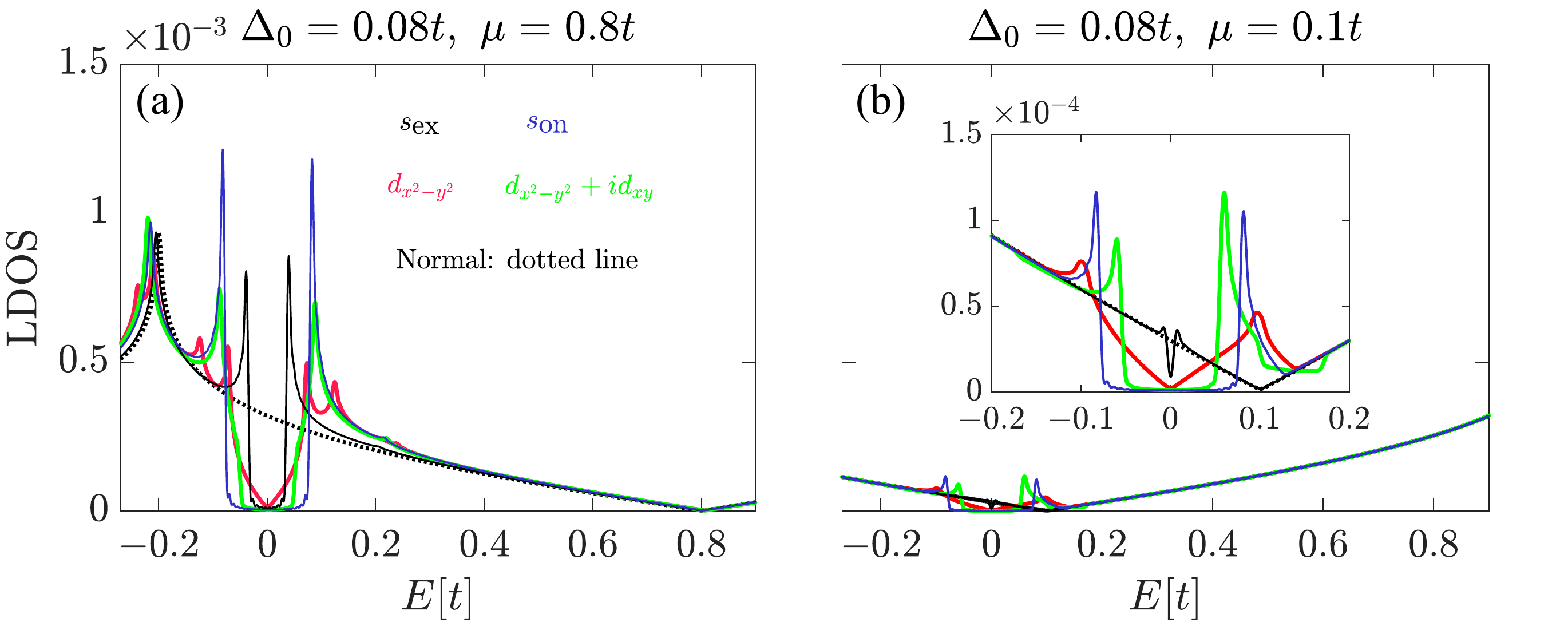}
	\caption[LDOS]{Bulk DOS of different states in graphene for $\mu=0.8t$, $\beta=10$ (a) and $\mu=0.1t$, $\beta=1.25$ (b). Inset in (b) is a zoom-in around at low energies.}
	\label{fig:LDOSU0M0801}
\end{figure}

As discussed in the main text, when $\beta=|\mu|/\Delta_0\gg 1$ the superconducting gap edge and $D_{g}$ are well separated and there is little or no interaction between superconductivity and the normal state Dirac point $D_{g}$. This is clearly shown in Fig.~\ref{fig:LDOSU0M0801}(a). In the bulk the $s$-waves are gapped with clear BCS-like coherence peaks.
The chiral $d$-wave is also fully gapped and although the gap respect the sixfold rotational symmetry of the lattice, it is not fully isotropic. The nodal $d$-wave symmetries, here represented by the $d_{x^2-y^2}$ state, have a Dirac point at zero energy, smaller but existing coherence peaks, and also some additional structures above gap edge, although the latter is not present for in the $d_{xy}$ state. All the superconducting states are thus Dirac point systems due to the normal state Dirac point at $E=\mu$. The nodal $d$-wave states have an additional Dirac point at $E = 0$ and are thus double Dirac point systems.

In the limit of $\beta \sim 1$, superconductivity is strong enough such that the superconducting coherence peaks covers the normal state Dirac point as illustrated in Fig.~\ref{fig:LDOSU0M0801}(b). Here $D_g$ is washed out resulting in  noticeable DOS at $E =\mu$. The exception is the $s_\text{ex}$ state, which does not exhibit this behavior because its gap reduces with doping and finally vanishes at half-filling.

\subsection{LDOS on impurity site}	
\begin{figure}[ht!]
	\centering
	\includegraphics[scale=0.440]{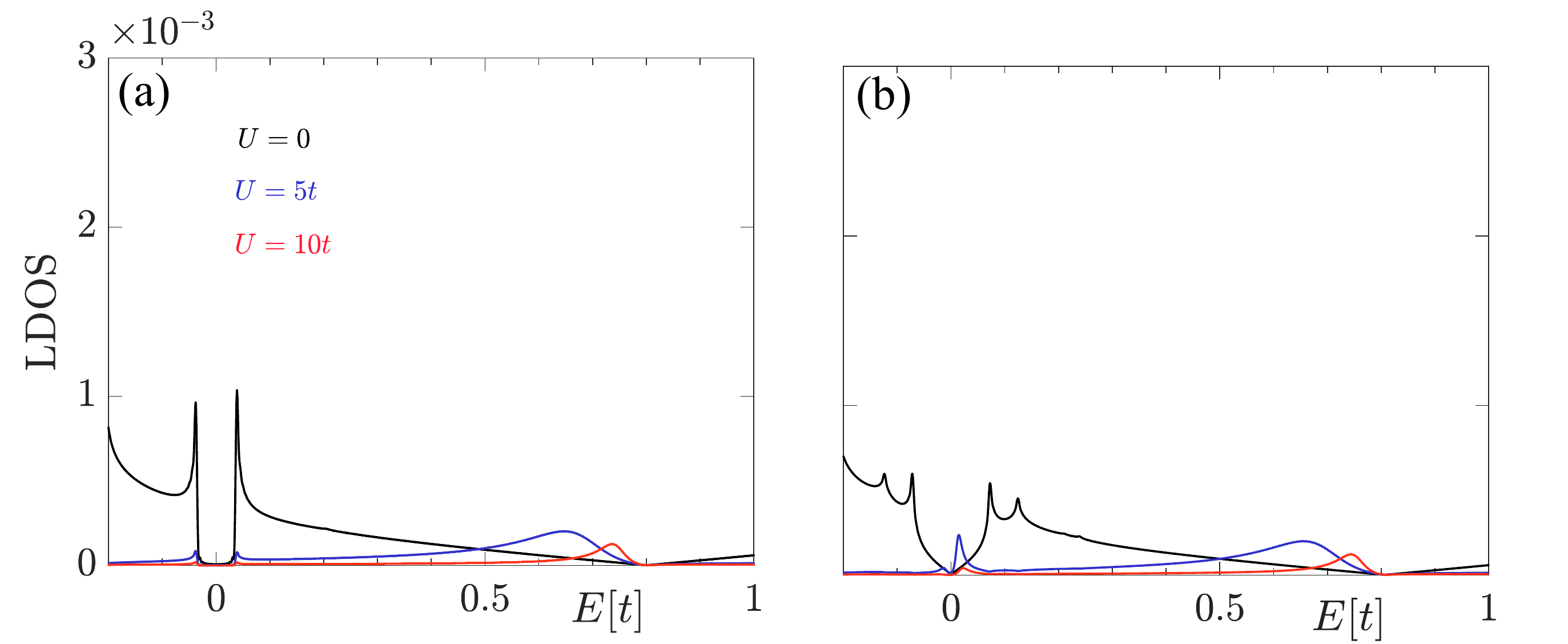}
	\caption[LDOS]{The equivalent figure of Fig.~2 (a,b) in the main text but for the LDOS at the impurity site for $s_\textrm{ex}$ (a) and $d_{x^2-y^2}$ (b) states.}
	\label{fig:LDOSImpSite}
\end{figure}
In the main text we stated that the impurity site has a strongly suppressed LDOS. This is due to the fact that hopping to the impurity site is suppressed by the impurity, with the higher values $U$ causing even more suppression. Figure~\ref{fig:LDOSImpSite} clearly illustrates this phenomena, where all impurity resonances are strongly suppressed. We therefore focus on impurity effects on NN sites (and beyond) of the impurity.

\subsection{Impurity resonance states at larger doping levels}
At large doping levels compared to the superconducting gap edge, such that $\beta \gg 1$, the normal state Dirac point $D_g$ and the superconducting gap are well separated in energy.
\begin{figure}[htb]
	\centering
	\includegraphics[scale=0.440]{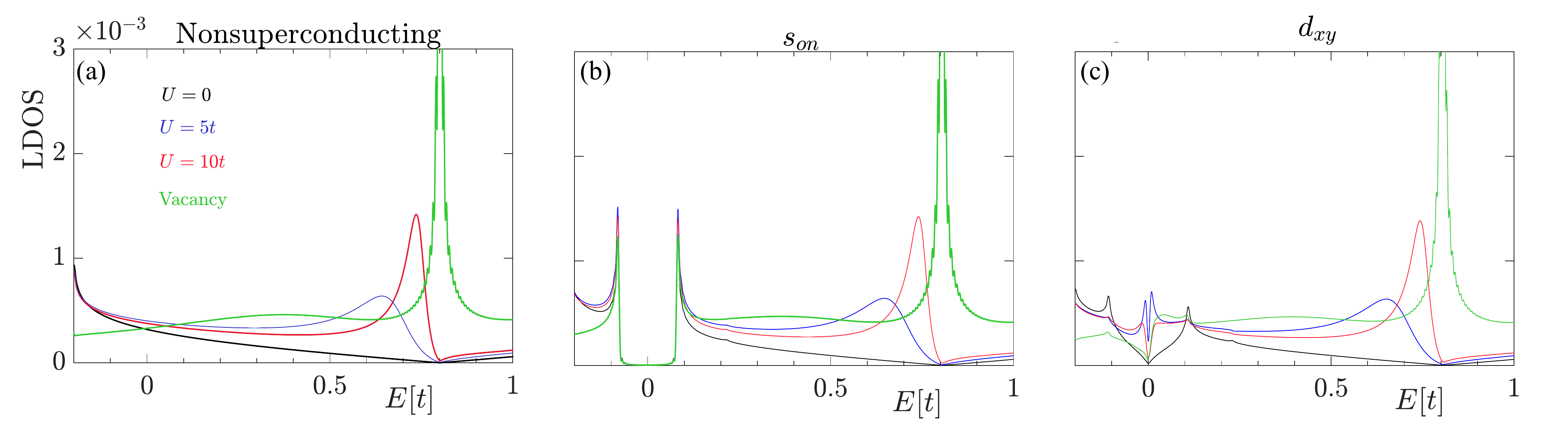}
	\caption[LDOS]{The equivalent figure of Fig.~2(a,b,c) in the main text but for the normal state (a) and $s_\textrm{on}$ (b) and  $d_{xy}$ (b) superconducting states.
	}
	\label{fig:LDOSMu08Usup}
\end{figure}
In the main text we stated that the resonance peak close to $D_g$ is due to the normal state. In Fig.~\ref{fig:LDOSMu08Usup}(a) we show exactly the same plot but in the non-superconducting phase to confirm this. We see that the same resonance peak present in all superconducting states is also present in non-superconducting graphene.
Figures~\ref{fig:LDOSMu08Usup}(b,c) further show that the $d_{xy}$ ($s$) state responds to a potential impurity in the same manner as the $d_{x^2-y^2}$ $(s_\text{ex})$ states plotted in Fig.~2 in the main text. This completes the series of NN impurity LDOS spectra for all relevant superconducting states at large $\beta$. The resonance peak at $D_g$ due to vacancy is similar to the case of hydrogenating a single graphene site \cite{lado2016unconventional}.

\subsection{Impurity resonance states at doping close to the gap edge}
\begin{figure}[htb]
	\centering
	\includegraphics[scale=0.440]{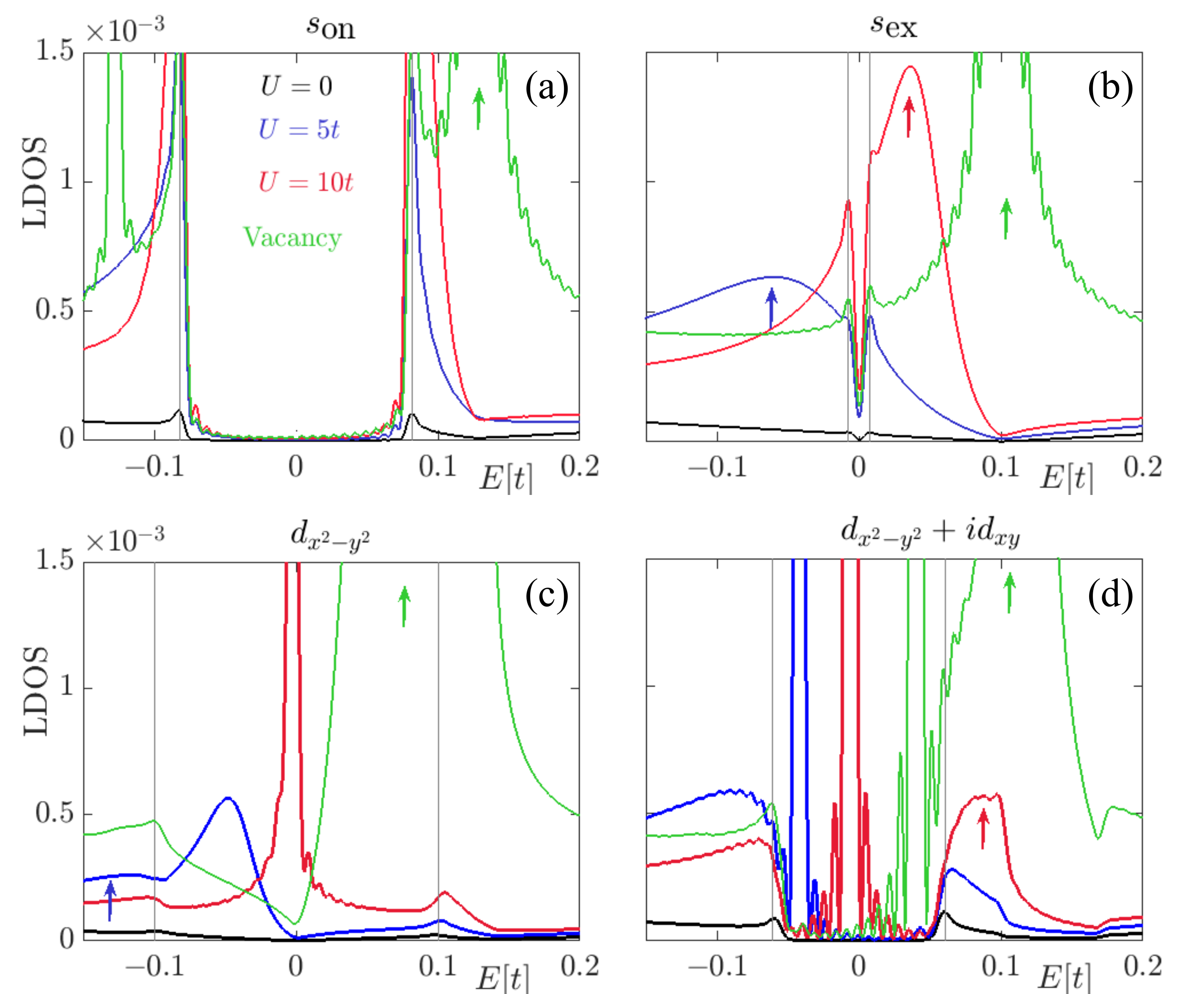}
	\caption[LDOS]{The equivalent figure of Fig.~2(a,b,c) in the main text but for $\mu=0.1t$ and $\Delta_0=0.08t$ giving $\beta=1.25$, and for $s_\textrm{on}$ (a), $s_\textrm{ex}$ (b), $d_{x^2-y^2}$ (c) and  $d_{x^2-y^2}+id_{xy}$ (d) superconducting states. Arrows indicate remaining identifiable supergap resonance peaks due to $D_g$, vertical grey lines  mark the bulk gap edges.}
	\label{fig:LDOSMu01U}
\end{figure}

As discussed in the main text, the behavior of the supergap resonance associated with the normal Dirac point $D_{g}$ changes when $\beta$ is reduced. This is illustrated in Fig.~\ref{fig:LDOSMu01U} which shows the results for $\beta = 1.25$ to be compared to Fig.~2 in the main text where $\beta = 10$. The supergap resonance is strongly suppressed and even largely disappears except in the unitary scattering limit. It is the suppression of the Dirac point by the superconducting coherence peaks that is responsible for this suppression. The exception is the $s_\text{ex}$ state in Fig.~\ref{fig:LDOSMu01U}(b), since here the Dirac point is not affected by superconductivity in the undoped case. We see further that there are no subgap resonances for either the $s$- or $s_{\rm ext}$-wave states. The $d$-waves, however, still exhibit clearly distingiushable subgap resonances. 

\subsection{Sign changes in impurity scattering and chemical potential}
In the main text we analyzed the situation with $\mu>0$ and $U>0$, i.e.~graphene is effectively hole-doped and the impurity scattering adds further to the on-site energy of the impurity site, such that it has on-site energy $\mu+U$. The impurity can thus be seen as a scatterer that will never counteract the overall doping of the system, just enhance it.
If we were to reverse the signs of both $U$ and $\mu$ simultaneously, all results are unchanged up to a reflection in energy. This situation can be understood as an electron doped system, where the impurity adds further electron doping. Thus, the electron doped  system has the same impurity resonances but with a reflection in energy: electron-like components are now at negative energies while the hole-like components have positive energies.

However, if only one of $U$ and $\mu$ changes sign, there is no subgap resonance states, rather the subgap features spread out in energy except in the unitarity limit. The supergap resonance still appears, but now at $E=2\mu-\text{sgn}(U)\Omega_g$. This different behavior is the result of the total on-site energy at the impurity site being a partial cancelation between the overall chemical potential and the impurity scatterer. A detailed understanding of the lack of subgap peaks  can relatively easily be obtained from the analytical solution of the poles of the $T$-matrix, but that is not the current objective.

\subsection{Additional results for graphene on cuprate substrate}
\begin{figure}[ht!]
	\centering
	\includegraphics[scale=0.440]{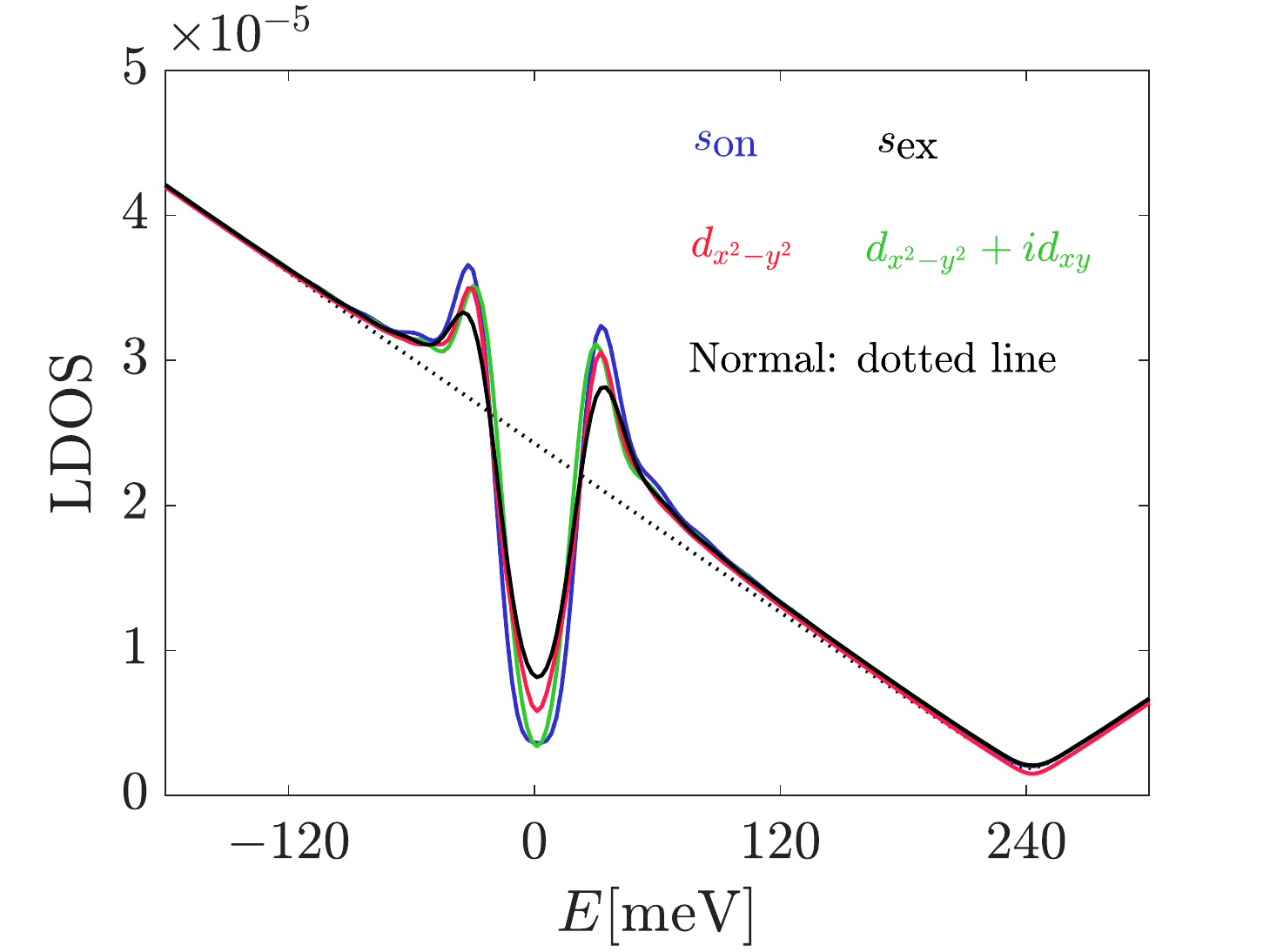}
	\caption[LDOS]{ Bulk DOS for different superconducting states. Here $\mu=0.08t=240$~meV and $\Delta_0=0.008t=24$~meV giving $\beta=10$, are the same as in Fig.~3 in the main text.}
	\label{fig:CleanMu008}
\end{figure}

We here present additional details relevant to experiments such as graphene deposited on a cuprate substrate. 
Using experimental parameters and in the absence of any impurities, all the superconducting states give very similar V-shaped LDOS profiles even with using very high resolution, see Fig.~\ref{fig:CleanMu008}. 
It is thus not possible to distinguish between different superconducting symmetries with bulk measurements alone, unlike the case with larger parameters shown in Fig.~\ref{fig:LDOSU0M0801}. The LDOS profiles in Fig.~\ref{fig:CleanMu008} are remarkably similar to that of the experimental STS results of graphene on a cuprate substrate in Ref.~[\onlinecite{di2017corrigendum}], with the only difference being the coherence peaks being a bit more washed out. 

\begin{figure}[ht!]
	\centering
	\includegraphics[scale=0.440]{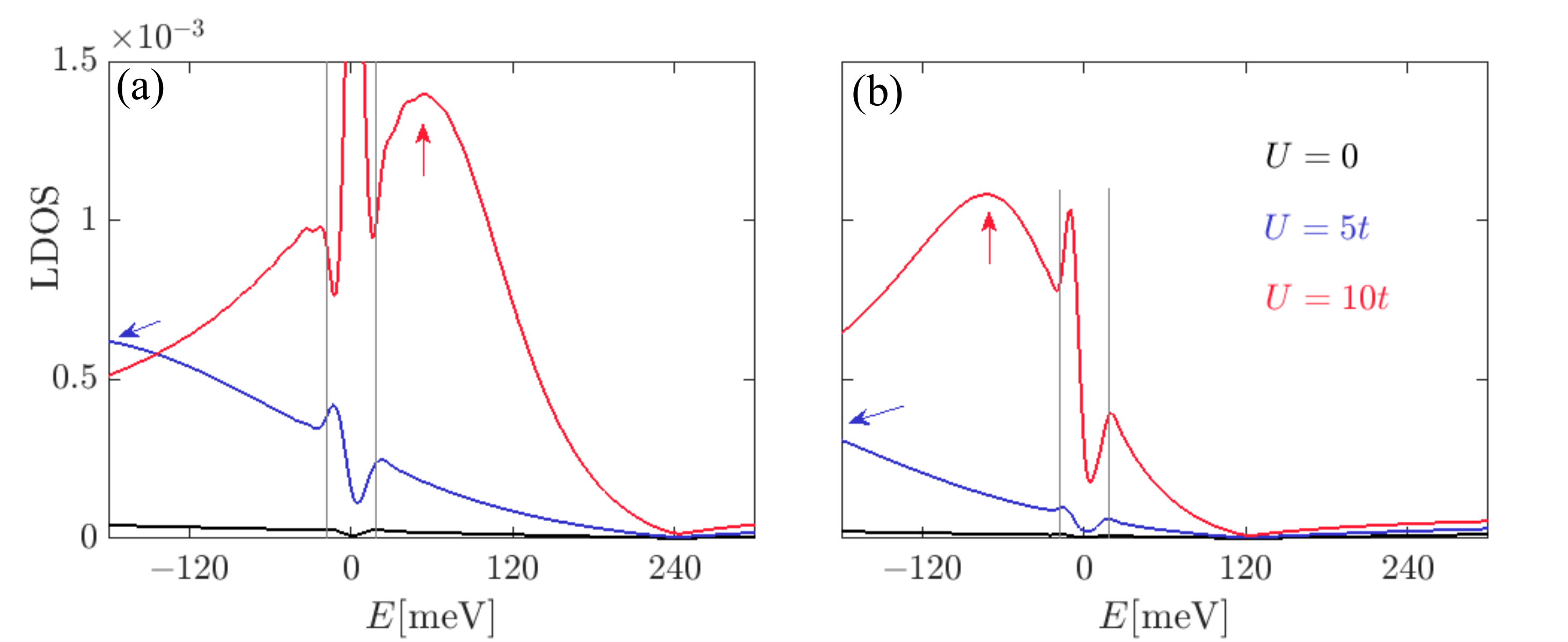}
	\caption[LDOS]{Equivalent figure to Fig.~3(b) in the main text for the  $d_{x^2-y^2}$ state but for $\Delta_0=0.004t=12$~meV and $\mu=0.08t=240$~meV  giving $\beta=20$ (a) and  $\mu=0.04t=120$~meV giving $\beta=10$ (b). Arrows indicate resonance peaks due to $D_g$, vertical grey lines mark bulk gap edges.}
	\label{fig:LDOSMu004}
\end{figure}

In Fig.~$3$ of the main text we showed how a potential impurity strongly differentiates between the different superconducting states, even when the bulk LDOS spectra are all very similar. Here we provide further supporting data with the gap reduced even further to $\Delta_0=0.004t=12$~meV. As shown in Fig.~\ref{fig:LDOSMu004}, the LDOS response to an impurity for the $d$-wave state is qualitatively the same as for the larger gap values in Figs.~2(b) and 3(b) in the main text. Further changing the doping level, compare Figs.~\ref{fig:LDOSMu004}(a) and (b), mainly repositions the supergap resonance. This is the same phenomena as seen in Fig.~3 in the main text and due to the supergap resonance position being primarily determined by the distance to $D_g$.

\subsection{Spatial LDOS pattern of supergap resonance}
\begin{figure}[ht!]
	\centering
	\includegraphics[scale=0.440]{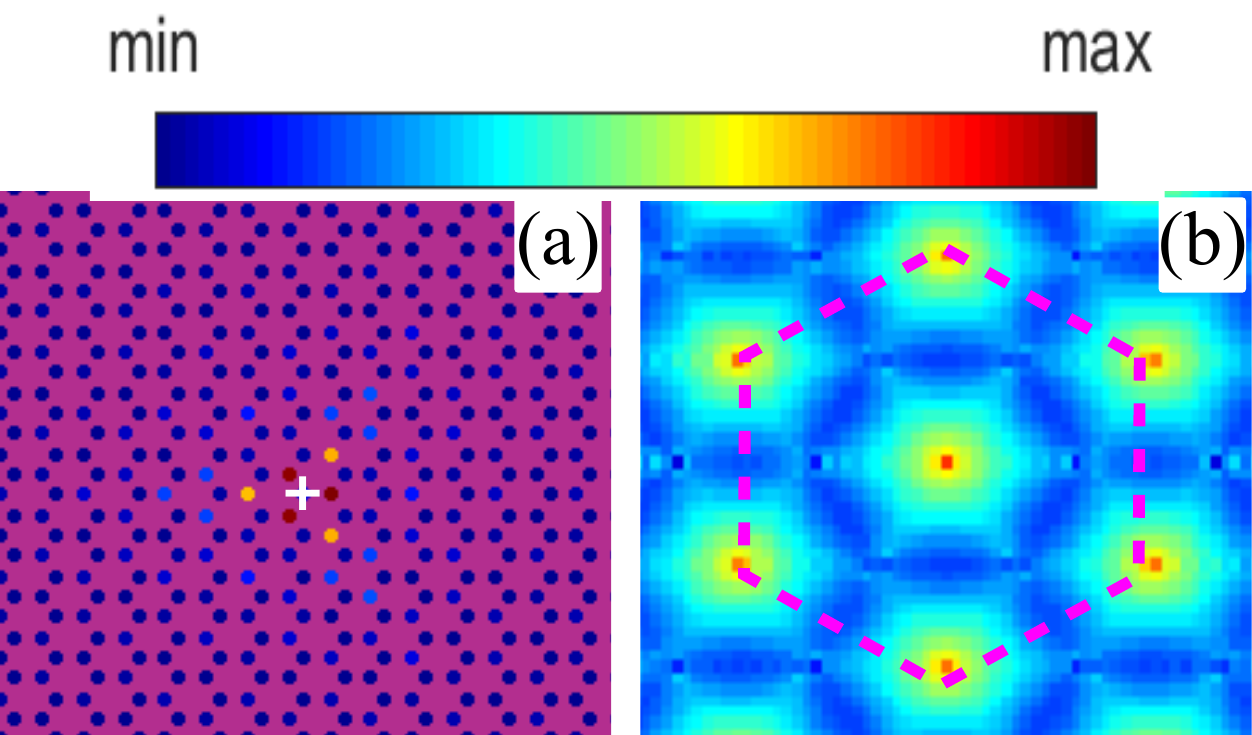}
	\caption[LDOS1]{The equivalent figure of Fig.~4 in the main text but for the (positive energy) supergap impurity resonance peak in the normal state.}
	\label{fig:spLDOSDg}
\end{figure}

In the main text we investigated the spatial properties of the impurity-induced LDOS for subgap resonances. Here we show the same data for the supergap resonance. In Fig.~\ref{fig:spLDOSDg} we plot the spatial extension of the impurity effect on both LDOS and FTLDOS obtained at supergap impurity peak in the normal state. It does not change when adding any superconducting term.
The real space impurity scattering LDOS pattern for normal graphene has been calculated in earlier studies~\cite{wehling2007local} and our results are similar, showing a very localized state obeying all lattice symmetries. The FTLDOS is notably different from the subgap resonance patterns, illustrating yet again how the supergap resonance is due to the normal state. We note that the QPI shown by the FTLDOS is qualitatively the same as obtained previously for graphene~\cite{bena2005quasiparticle,rutter2007scattering}.

\subsection{Spatial LDOS pattern of subgap resonances}
\begin{figure}[ht!]
	\centering
	\includegraphics[scale=0.440]{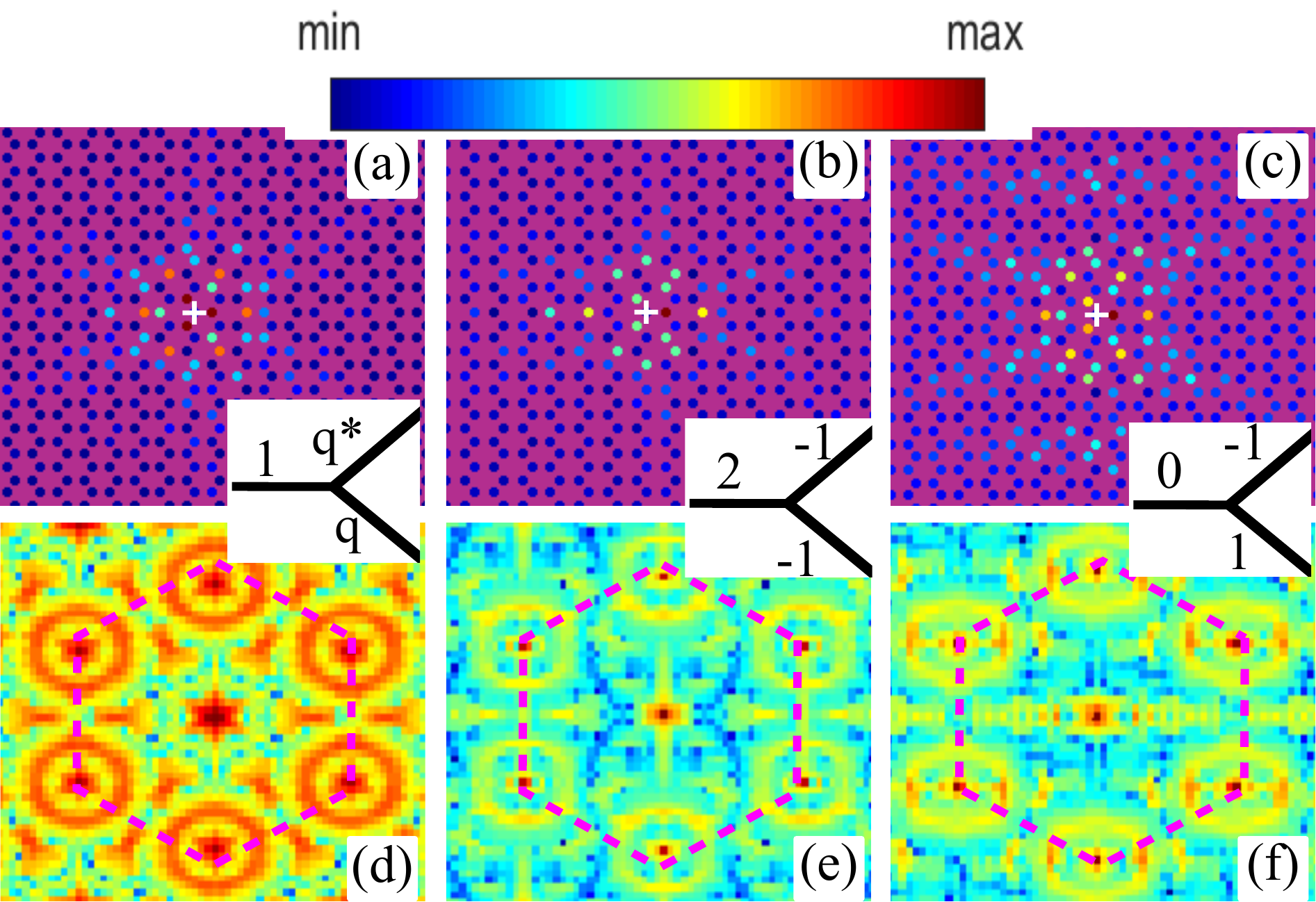}
	\caption[LDOS]{The equivalent figure of Fig.~4 in the main text but for positive subgap impurity resonance peak energies.}
	\label{fig:spLDOSEp}
\end{figure}

In the main text we studied only the LDOS spatial pattern of the negative energy subgap impurity resonance. In Fig.~\ref{fig:spLDOSEp} we provide the complementary results for the positive energy peak.
The conclusions are similar in that the chiral $d$-wave state has both LDOS and FTLDOS patterns respecting the sixfold lattice rotational symmetry, while the nodal $d$-wave states has only left a twofold rotational symmetry. There is however a sort of complementariness between Fig.~\ref{fig:spLDOSEp} taken at positive energy and Fig.~4 in the main text at the same but negative energy. This is to be expected as on each site in the lattice, the sum of the magnitude of the hole and electron components of the quasiparticles is always unity \cite{Balatskyet.alRMP2006}.

\begin{figure}[ht!]
	\centering
	\includegraphics[scale=0.430]{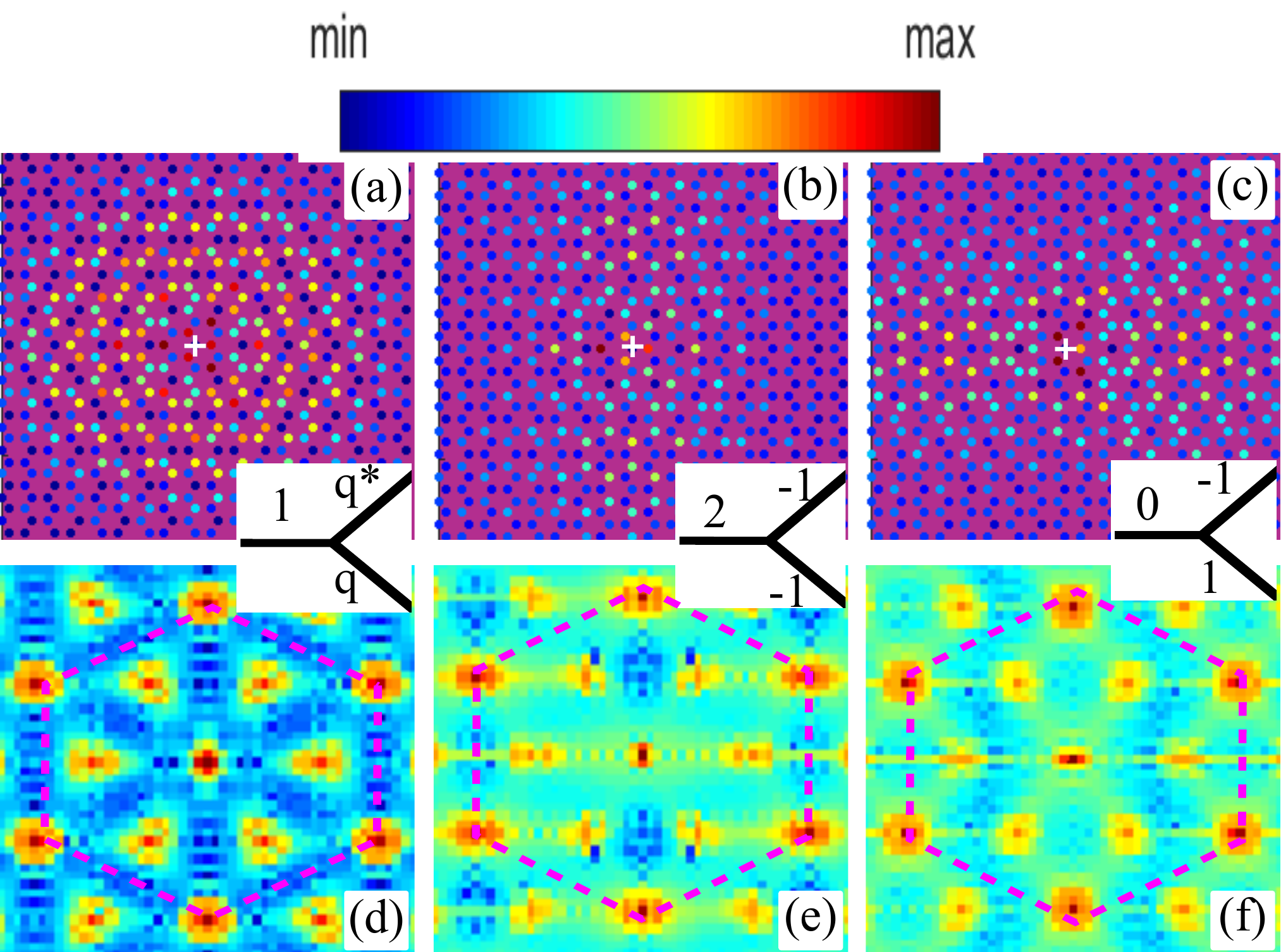}
	\caption[LDOS]{The equivalent figure of Fig.~4 in the main text but for smaller $\mu = 0.3t$.}
	\label{fig:spLDOSEnMu3}
\end{figure}
We also plot in Fig.~\ref{fig:spLDOSEnMu3} the LDOS and FTLDOS for the negative energy subgap impurity resonances at a lower doping level $\mu=0.3t$. This results in a reduced $\beta=3.75$ compared to $\beta = 10$ in Fig.~4 in the main text. In this case, the normal state Dirac point $D_g$ is much closer to the superconducting gap edge and the impurity state spreads out more in space for all superconducting states. Also, the nodal $d$-waves has a more one-dimensional spread compared to at larger doping levels. It is therefore an even clearer difference between the nodal and chiral $d$-wave states in this parameter regime. Overall, the rings are also smaller in the FTLDOS data since the Fermi surface shrinks with doping level. 
Finally, for doping extremely very close to the charge neutrality point and small $\Delta_0$, the spatial patterns are similar to that of normal graphene. This is because graphene does not support superconductivity at $\mu  \rightarrow 0$ due to vanishing density of states close to the $K$ and $K'$ points of the BZ.

\end{document}